\documentclass[conference]{IEEEtran}

\ifCLASSINFOpdf
\else
\fi

\usepackage{multirow}
\usepackage{siunitx}

\usepackage{graphicx}
\usepackage{adjustbox}
\usepackage{multirow}
\usepackage[table,xcdraw]{xcolor}
\usepackage{xcolor}
\definecolor{redcolor}{rgb}{1, 0, 0}

\begin{document}
%
\title{An Automated Framework for Board-level \\Trojan Benchmarking}
\author{\IEEEauthorblockN{Tamzidul Hoque, Shuo Yang, Aritra Bhattacharyay, Jonathan Cruz, and Swarup Bhunia }
\IEEEauthorblockA{Department of Electrical and Computer Engineering\\
University of Florida,
Gainesville, Florida 32611\\
Email: \{thoque, sy, abhattacharyay, jonc205\}@ufl.edu, swarup@ece.ufl.edu}
\vspace{-7 mm}

%
%
%
%

}

%


\maketitle

\begin{abstract}
Economic and operational advantages have led the supply chain of printed circuit boards (PCBs) to incorporate various untrusted entities.
Any of the untrusted entities are capable of introducing malicious alterations to facilitate a functional failure or leakage of secret information during field operation.
While researchers have been investigating the threat of malicious modification within the scale of individual microelectronic components, the possibility of a board-level malicious manipulation has essentially been unexplored. In the absence of standard benchmarking solutions, prospective countermeasures for PCB trust assurance are likely to utilize homegrown representation of the attacks that undermines their evaluation and does not provide scope for comparison with other techniques. In this paper, we have developed the first-ever benchmarking solution to facilitate an unbiased and comparable evaluation of countermeasures applicable to PCB trust assurance. Based on a taxonomy tailored for PCB-level alterations, we have developed high-level Trojan models. From these models, we have generated a custom pool of board-level Trojan designs of varied complexity and functionality. We have also developed a tool-flow for automatically inserting these Trojans into various PCB designs and generate the Trojan benchmarks (i.e., PCB designs with Trojan). The tool-based Trojan insertion facilitate a comprehensive evaluation against large number of diverse Trojan implementations and application of data mining for trust verification. Finally, with experimental measurements from a fabricated PCB, we analyze the stealthiness of the Trojan designs.

\end{abstract}

\begin{IEEEkeywords}
Printed Circuit Board (PCB) , Trust Verification, Hardware Trojan 
\end{IEEEkeywords}

%
\IEEEpeerreviewmaketitle

\section{Introduction}
Printed circuit boards (PCBs) are the backbone of almost all electronic systems. They contain a diverse set of microelectronic components starting from simple passive devices like diodes, transistors, resistors, capacitors to complex integrated circuits (ICs) like microprocessors, field-programmable gate arrays (FPGA), and memories. The design and fabrication process of these PCBs involves several untrusted vendors capable of introducing malicious change in the system to facilitate a detrimental outcome in the field. Such inclusion of PCB-level malicious circuits at foreign facilities has already been reported with the affirmation from 17 anonymous private and government agency sources \cite{bighack}. According to the report, computer server motherboards used in at least 30 US corporate and government entities have been found compromised with this supply chain attack. Even though many have questioned the authenticity of this attack, the article, nevertheless, highlights board-level malicious inclusion as a serious threat to the trustworthiness of modern electronic system.  This threat calls for effective countermeasures that can prevent, detect, and tolerate such attacks. While a vast amount of research has explored the various facets of malicious circuits inside individual microelectronic components, a more complex attack vector that introduces board-level malicious functionalities has received little to no attention \cite{xiao2016hardware}.   

Modeling the attack vector is a critical step towards assuring trust at the PCB-level. The construction and behavior of malicious circuits can be completely different depending on which entities in the PCB supply chain we assume to be untrusted. In the absence of a comprehensive taxonomy of PCB-level Trojans and associated benchmarks, research in the domain of PCB trust can suffer from the following challenges:    
\begin{itemize}
    \item Researchers are likely to validate their countermeasures using hand-crafted Trojan instances. Such home-grown designs are not guaranteed to meet the design requirements that allow the Trojans to go undetected under existing inspection and verification techniques.  
    \item Validating the countermeasures using hand-crafted malicious circuits that are publicly unavailable deters the scope for comparing different solutions. 
    
\end{itemize}

On the other hand, a suite of publicly available custom Trojan benchmarks with few representative examples for different classes of Trojans have the following deficiencies:   
\begin{itemize}
    \item Custom Trojan benchmarks only represents a small subset of the possible design variations an attacker could explore. While certain countermeasures may work well for those limited number of Trojans due to some inherent bias in their designs, the solutions may still fail when applied in the field.   
    \item Application of supervised machine learning based solutions in detecting Trojans require a large number of examples to train from \cite{hoque2018hardware}. Training data obtained from the limited examples available in a custom Trojan benchmark suite may not be adequate for generating a robust trained model capable of identifying the intended malicious behaviour.  
  
\end{itemize}

To eliminate the aforesaid challenges, we have developed a detailed taxonomy of PCB-level malicious circuits with various examples of custom Trojans in the form of schematic and layout that are suitable for static analysis, dynamic analysis, and fabrication. We also have developed the first-ever tool, to the best of our knowledge, to automatically create PCB benchmarks by inserting these Trojans in various PCB designs. 

The Trojan benchmark suite is built upon 10 open-source baseline PCB designs of various complexities.
Each of these baseline designs are inserted with different types of Trojans to generate around 150 infected PCB designs. The baseline designs are of different complexities and sizes. 
To define the Trojan models, we first develop a dedicated taxonomy of PCB-level Trojans. The taxonomy classifies the design space of board-level Trojans with respect to their phase of insertion in the supply chain, abstraction level of the PCB, activation mechanism, the impact of payload, location of insertion, and physical characteristics. Compared to existing taxonomy of IC-level Trojans \cite{shakya2017benchmarking}, PCB-level Trojans differ in various aspects. For instance, while the integration of Trojan after fabrication is considered infeasible at the IC-level, it is a potential threat for PCBs as new components can be soldered onto the board even after fabrication.

Finally, to insert a large number of diverse Trojans in different PCB designs, we have developed a Trojan insertion tool. The tool takes the template of a Trojan model and a sample PCB design in the form of a netlist. 
It also takes some configuration information as inputs to define the number of Trojans and type of Trojan and then inserts them into the sample design. 
Overall, we make the following major contributions:
\begin{itemize}
    \item We rework the existing classification of malicious circuits to develop a taxonomy of hardware Trojan attacks applicable to PCB. 
    \item We develop several custom Trojan designs for PCB with different trigger and payload mechanisms. 
     \item We present a toolflow to automatically insert a large number of Trojans in PCB designs. Such a tool is beneficial for a comprehensive evaluation of trust verification techniques and application of machine learning.
    \item We evaluate the Trojan designs using experimental results obtained from a fabricated PCB containing several board-level Trojans.

\end{itemize}
 The rest of the paper is organized as the following: Section \ref{sec:background} describes the necessary definitions and existing studies on Trojan benchmarking. Section \ref{sec:custom} presents the Trojan taxonomy and the Trojan designs. Section \ref{sec:synthetic} elaborates the automatic Trojan insertion process. Experimental results are presented in Section \ref{sec:result} with conclusive remarks in Section \ref{sec:conclusion}.     
\section{Background}
\label{sec:background}
\subsection{Definitions}
\subsubsection{PCB-level Trojan}
PCBs consist of various microelectronic components spread across multiple alternating layers of conducting and insulating materials. Conductive planes across different layers are connected through vias. 
A PCB-level Trojan can be designed by introducing additional components that do not contain any malicious sub-circuit internally. For instance, to construct a PCB-level Trojan, an attacker may introduce several additional NPN transistors that strictly follow the behaviour of a generic NPN transistor but do not have any relation in serving the intended functionality of the PCB.    
Board-level Trojans could also be introduced without integrating additional components by only altering the structural and/or parametric behaviour of conducting (e.g., copper traces, vias, etc.) and insulating parts of the PCB layers. 
More complex attack vectors utilizing both intra-board and intra-component modification are possible as well. However, in this paper, we assume that the individual microelectronic components performs their designated functionality and there is no malicious functionality hidden \textit{inside} the components. 
\subsubsection{PCB Trojan Benchmarks}
A PCB Trojan benchmark is a PCB design with one or more PCB-level Trojan modifications. The benchmark could be in the form of specification, netlist, schematic, or layout.   

\subsection{Related Work}
Existing work on PCB security primarily involves counterfeit PCB detection \cite{hennessy2016jtag, zhang2015robust} and detection of PCB tampering in the field \cite{paley2016active}. 
The possibility of hardware Trojan attacks in PCB was first discussed in \cite{ghosh2014secure} with a general models of Trojans attacks applicable to PCB. They also demonstrate two specific attack instances assuming different sets of supply chain entities to be untrusted. However, it does not provide a detailed taxonomy for PCB-level Trojans and does not present any benchmark suite.     
Existing work on hardware Trojan taxonomoy and benchmarking covers malicious modification of digital IC designs \cite{shakya2017benchmarking}. A benchmark suite containing 96 Trojan inserted designs are available in the Trust-HUB website. These designs are provided as register-transfer-level code, gate-level netlist, and layout. The Trojans are inserted in various IP cores such as AES, RS232, MP8051, etc. The suite contain few representative Trojan benchmarks from various broad classes of Trojans in the taxonomy. 

 To eliminate the limitations inherent to a static benchmark suite, \cite{cruz2018automated} proposed a tool-based Trojan insertion process to generate a diverse possible implementation for each class of Trojans. The tool takes a sample gate-level design of an IC and some configuration parameters that define the functional and structural behaviour of the Trojans to be inserted. The tool identifies potential regions in the sample design for Trojan insertion and integrates a gate-level Trojan circuit. It also verifies if the Trojan can be activated under specific input pattern to the design. Similar to Trust-HUB Trojans, the current tool-based Trojan insertion methods do not serve the benchmarking purpose for PCB-level Trojan attacks.
 \begin{figure*}[t]
 	\centering
 	  \includegraphics[width=\textwidth]{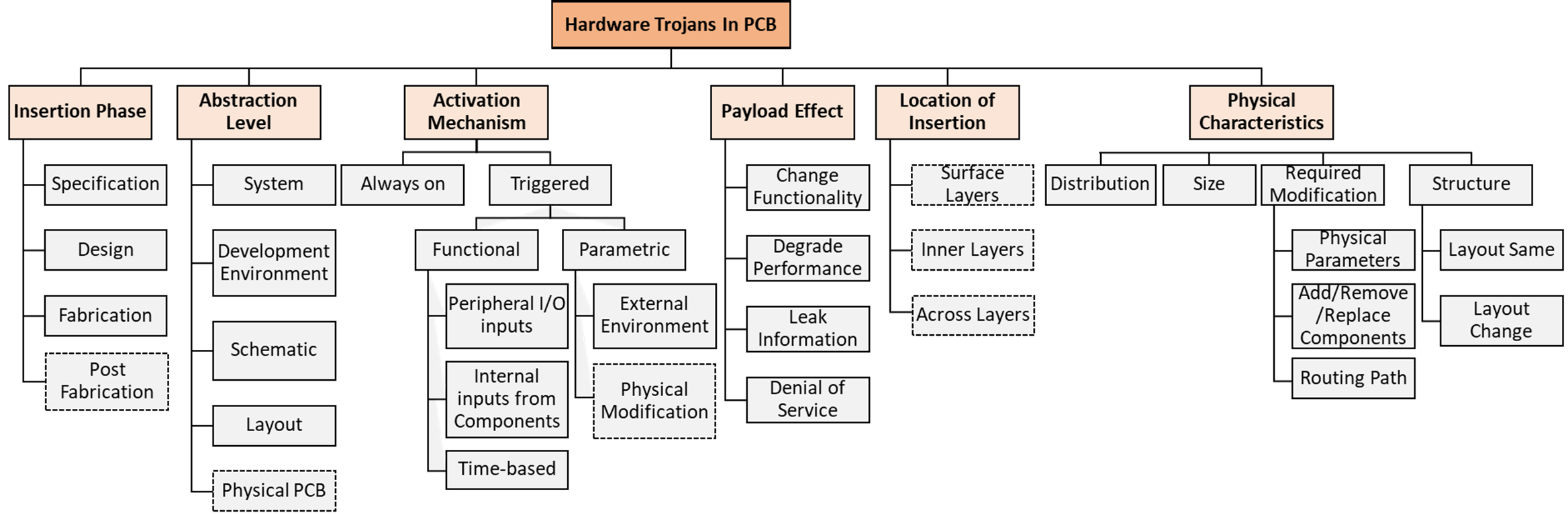}
 	\caption{Taxonomy of PCB-level hardware Trojans. The dotted-box show the Trojan classes primarily applicable to PCBs.}
 	\label{taxo}

 \end{figure*}

 \begin{figure}[t]
 	\centering
 	  \includegraphics[width=.9\columnwidth]{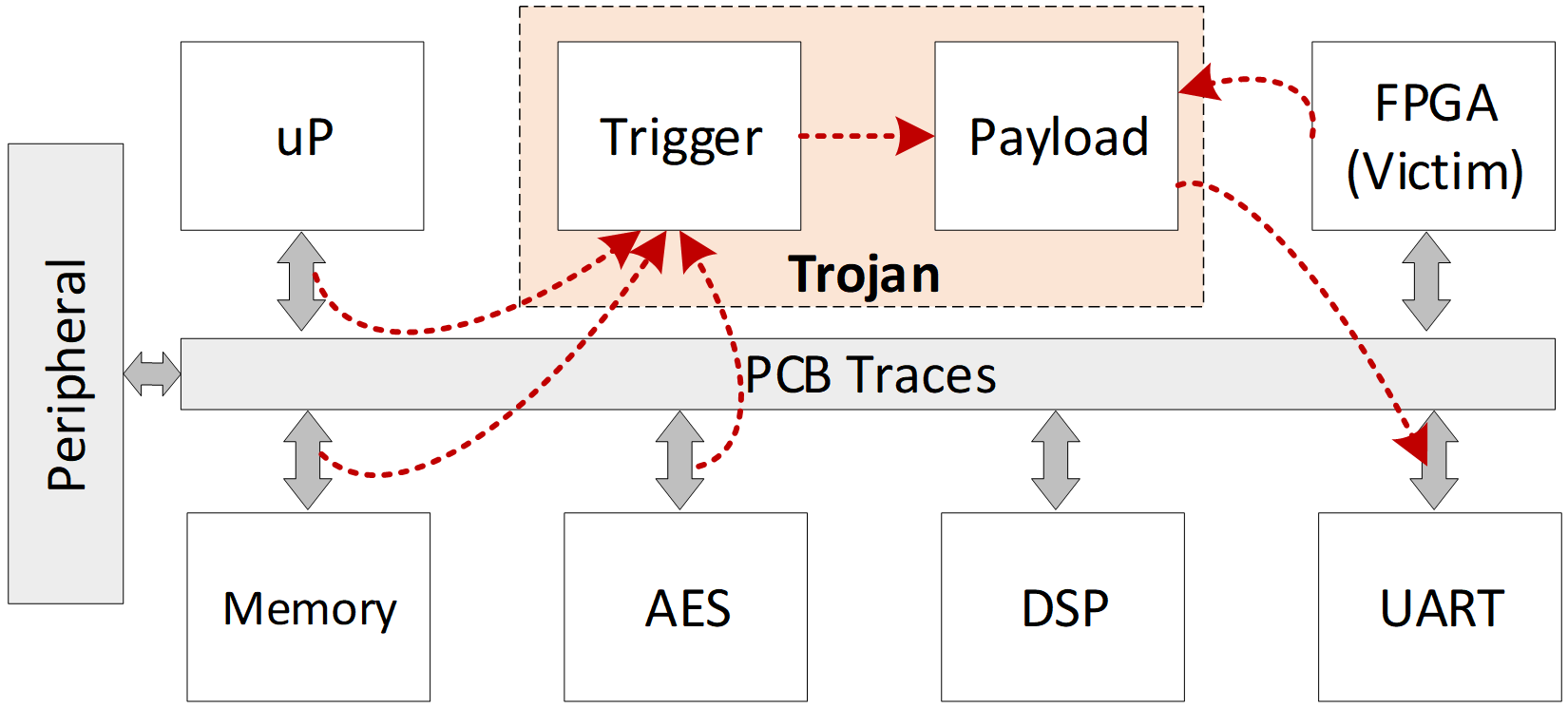}
 	\caption{High-level illustration of the \textit{Addition Model}. We present various methods to construct hard-to-activate trigger circuits and payload circuits to leak a victim signal or corrupt its functionality.  }
 	\label{fig:model}

 \end{figure}
\section{Hardware Trojan Attack in PCB}
\label{sec:custom}

\subsection{PCB Trojan Taxonomy}
The only existing classification of PCB-level Trojans divides the Trojan space into two broad categories~\cite{ghosh2014secure}. In the first category, the the design house is assumed to be trusted, and in the second one, both design house and foundry are considered to be untrusted. To precisely identify the coverage, strengths, and weaknesses of various countermeasures, a taxonomy of Trojans with fine-grained classification is required. Again, existing taxonomy for IC-level Trojans do not directly apply to PCB as their design abstractions, flexibility for modification, and post-fabrication physical access to the internals are different \cite{shakya2017benchmarking}. Hence, based on the existing studies, we have developed a fine-grained classification for PCB-level Trojans as presented in Fig. \ref{taxo}. We can observe several categories of Trojans that are only applicable for PCBs (marked with dotted box). For instance, addition of new components or alteration of traces after fabrication and assembly of the board is feasible in PCB \cite{paley2016active}. Such modifications are usually considered as invasive physical attacks. The goal of PCB physical tampering is to observe an immediate malicious outcome during the attack, such as modchip attacks in gaming consoles \cite{fitzgerald2005playstation}. However, Trojan insertion, whether before or after fabrication, is likely to occur with a goal to compromise the system in a later stage of deployment. 
Most in-field invasive attacks on ICs require significant resources for decapsulation, delayering, and modification at the nano-scale level \cite{rahman2018physical}. For PCBs, physical alteration after fabrication is more feasible, which motivates us to consider post-fabrication alteration in the taxonomy. Physical alteration could also be used as an activation mechanism for certain Trojans. For instance, thinning of traces at the surface layer is very feasible and can be used to trigger a failure from heating generated during long hours of operation \cite{ghosh2014secure}. The location of Trojan insertion in a PCB can be classified with respect to the target layer of alteration. For instance, PCB designs developed by third-parties may have malicious circuitry at the surface level of the PCB, as the consumers do not have a reference design. However, in the presence of a reference or ``golden'' design, the attacker is likely to target alterations within the internal layers to remain stealthy.


\begin{figure}[t]
 	\centering
 	  \includegraphics[width=\columnwidth]{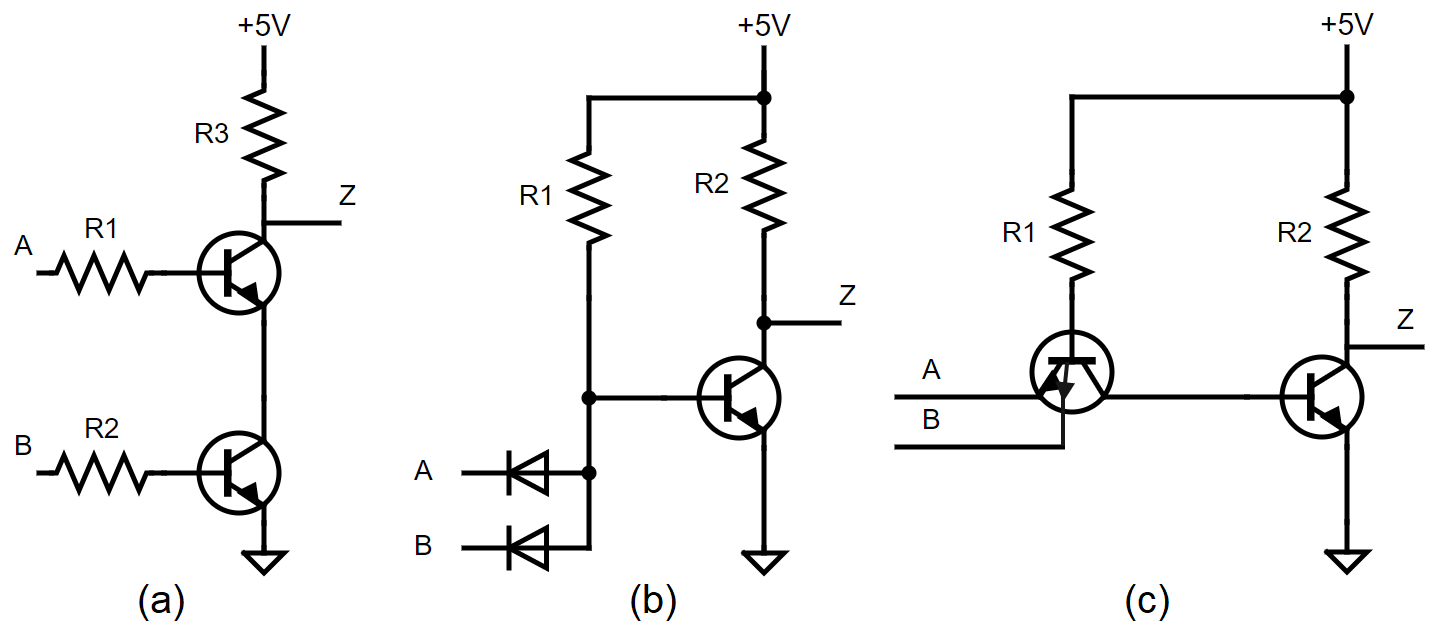}
 	\caption{Three different implementation of 2 input NAND: (a) using resistor-transistor logic, (b) diode-transistor logic, and (c) transistor-transistor logic.}
 	\label{fig:logic_family}

 \end{figure}
 \begin{figure*}[!t]
 	\centering
 	  \includegraphics[width=.98\textwidth]{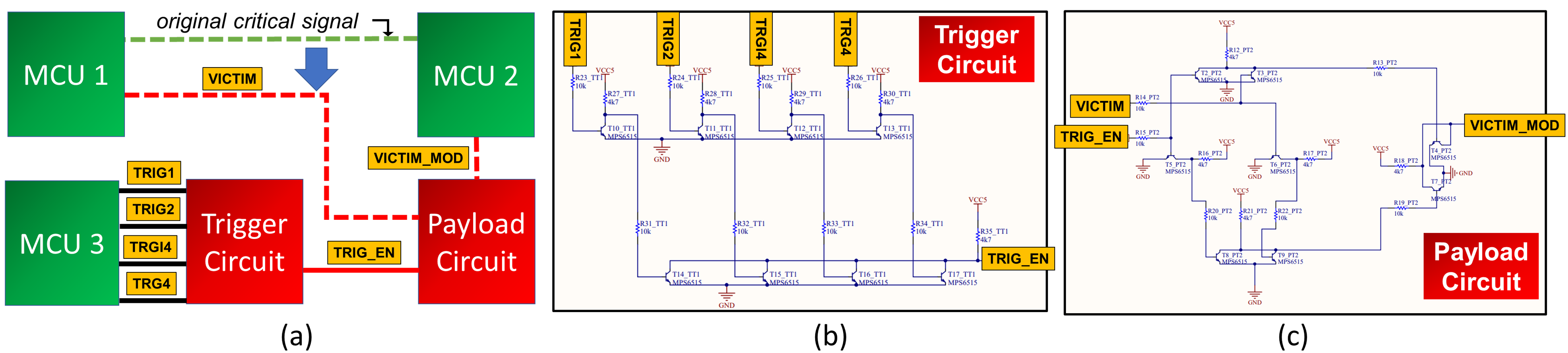}
 	\caption{A Trojan implanted with RTL logic: (a) a signal from from MCU 1 to MCU 2 (in green) is routed through a Trojan payload circuit (in red); (b) a trigger circuit (4-input AND function) activates the payload when input from MCU 3 is 4’b1111; (c) upon activation, the XOR payload inverts the victim.}
 	\label{fig:rtl}
 \end{figure*}
\subsection{Trojan Models and Designs}
We use two high-level Trojan models to generate several Trojan designs. These designs represent various Trojan classes within the taxonomy. The first model requires the addition of new components, thus named \textit{Addition Model}. The second model is called \textit{Alteration Model} that alters the existing PCB layout without adding new components.

 \subsubsection{Addition Model}
Fig. \ref{fig:model} shows the generic structure of this model that contains a trigger and a payload. The trigger takes one or more inputs from the existing traces of the board to construct a hard-to-activate trigger condition. Rare activation is needed to minimize the probability that the Trojan gets accidentally triggered during functional testing of the board. Difficult activation can be ensured by integrating large number of traces that rarely reach to a specific activation condition, or by constructing a trigger circuit that requires multiple excitations of a condition. The payload can be designed to achieve a number of malicious outcomes as described in the taxonomy. Certain implementations may only contain a payload (i.e., always-on Trojan \cite{lin2009moles}). However, such Trojans must not cause any observable impact on the functionality of the board as they are trivial to detect.

 For the addition model, we assume that the PCB design is untrusted. Hence, the consumer has no reference to identify components that are included for malicious purposes. However, when selecting components for constructing the Trojan, we try to utilize components that are very common in a PCB design. For instance, diode, NPN/PNP transistors, resistor, capacitors, op-amps are commonly found in a PCB designs. Therefore, their presence in a third-party PCB design is unlikely to raise question, specially if the board originally contained them to provide the normal functionality. Including a standalone microprocessor solely to perform a malicious operation may allow easier identification through disassembly and analysis of the program  being executed on it. Below we discuss various trigger and payload designs in this model.

\begin{figure}[t]
 	\centering
 	  \includegraphics[width=\columnwidth]{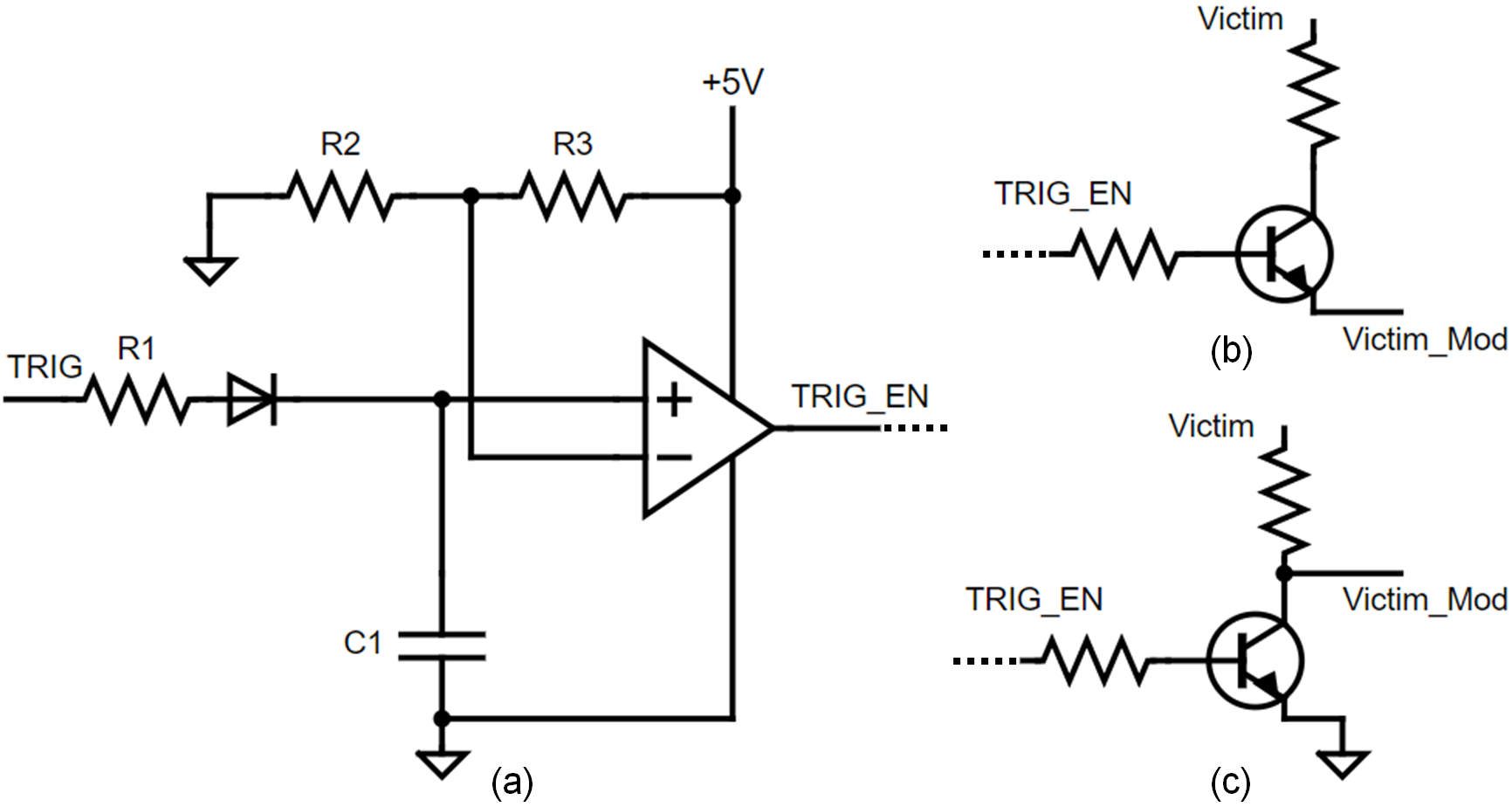}
 	\caption{(a) Trigger circuit similar to A2 Trojan \cite{yang2016a2} where capacitor and op-amps are used to design activation mechanism that requires the rare condition to occur multiple times. Payload with single transistor: (b) impacts the victim when the transistor is off, (c) impacts the victim when the transistor is on.}
 	\label{fig:A2}
 \end{figure}

\textbf{Example Trigger Designs:}
To construct hard-to-activate trigger circuits using diode, NPN/PNP transistors, resistor, capacitors, and op-amps, we look at various logic families including Resistor-Transistor Logic (RTL), Diode-Transistor Logic (DTL), and Transistor–Transistor Logic (TTL) \cite{tietze2008logic}. These logic technologies can be used to implement any type of Boolean function. As shown in Fig. \ref{fig:logic_family}, a 2-input NAND function can be constructed differently using RTL, DTL, and TTL design methods. This also provides a way to generate a structural variant of a given trigger functionality. Fig. \ref{fig:rtl} (a) shows one example of a Trojan that could be integrated by an untrusted design house to alter the state of a critical signal sent from MCU 1 to MCU 2. Fig. \ref{fig:rtl} (b) shows the corresponding trigger logic implementing a 4-input AND function in RTL method. When the four inputs of the trigger circuit (i.e., \texttt{TRIG1} to \texttt{TRIG4}) become logic high simultaneously, the output of the trigger logic (\texttt{TRIG\_EN}) becomes logic high. We can construct a trigger logic cone of any number of inputs using this method. To incorporate a counter-based trigger condition, we take inspiration from the Trojan design methods presented in A2 \cite{yang2016a2}. As shown in Fig. \ref{fig:A2}, using a capacitor, diode, and Op-Amp, it is possible to construct a trigger circuit that only activates when the capacitor is charged to reach a reference voltage defined by the two feedback resistors connected to the Op-Amp. Each time \texttt{TRIG} stays high, the capacitor starts charging and the charge stays in the capacitor even if \texttt{TRIG} goes low. Hence, if we remove the diode, the trigger condition becomes even harder to achieve due to the discharge path for the capacitor. In that case, to make \texttt{TRIG\_EN} high \texttt{TRIG} must maintain high for a specific amount of time until the capacitor gets charged completely.

 \begin{table*}[!t]
 \caption{Sample PCB Designs for Inserting Trojans}
 \label{tab:sample}
\scalebox{.95}{
\begin{tabular}{|l|l|l|c|c|l|}

\hline
\# & Name of the board & Dimension         & No. of Components & No. of Layers & Example Components                                                                                                     \\ \hline
1  & CIAA-ACC          & 95.25mm*91.44mm    & 569               & 12           & R, C, D, Sw,  $\mu C$, FPGA, DRAM, Transistor,  Ethernet,  JTAG                    \\ \hline
2  & EDU\_CIAA\_K60    & 76.4mm*137mm       & 187               & 2            & R, C, D, Sw,  $\mu C$, EEPROM,  Transreceiver, Transistor                                \\ \hline
3  & EDU\_CIAA\_Intel  & 90.17mm*80.01mm    & 87                & 2            & R, C, D, Sw,  $\mu C$, Transistor                                                              \\ \hline
4  & EDU\_CIAA\_NXP    & 76.4mm*137mm       & 196               & 2            & R, C, D, Sw, $\mu C$,  EEPROM, Transreceiver                                                   \\ \hline
5  & CIAA\_K60         & 137.008mm*86.03mm  & 467               & 4            & R, C, D, Sw,  $\mu C$, Transistor, OpAmp, Transreceiver,  JTAG, SPI        \\ \hline
6  & CIAA\_FSL\_Mini   & 71.12mm*99.695mm   & 142               & 4            & R, C, D, Sw,  $\mu C$,  Transreceiver, Transistor                                 \\ \hline
7  & CIAA\_NXP         & 138.43mm*86.36mm   & 443               & 4            & R, C, D, Sw,  $\mu C$, Ethernet, SPI, Transreceiver, OpAmp, BJT \\ \hline
8  & CIAA\_PIC         & 138.725mm*86.175mm & 435               & 4            & R, C, D, Sw,  $\mu C$, SPI, Oscillator, Transistors, SRAM, OpAmp         \\ \hline
9  & CIAA\_PICO        & 50.80mm*30.48mm    & 67                & 2            & R, C, LED,  D, $\mu C$, Transistor                                                            \\ \hline
10 & CIAA\_RX          & 135.896mm*83.948mm & 450               & 4            & R, C, D, Sw,  $\mu C$,  SPI, Transreceiver, DRAM, Transistor, OpAmp     \\ \hline
11 & CIAA\_SAFETY\_VTI & 95.890mm*90.170    & 283               & 2            & R, C, D, Sw,  $\mu C$, JTAG, Transreceiver, Flash memory, Transistor                           \\ \hline
12 & CIAA\_Z3R0        & 18.8mm*51.81mm     & 34                & 2            & R, C, LED,  D, $\mu C$                                                                         \\ \hline
\multicolumn{6}{|l|}{\textit{*R: Resistor, C: Capacitor, D: Diode, Sw: Switch, $\mu C$: microcontroller}}                                                          \\ \hline
\end{tabular}}
\end{table*}

\textbf{Example Payload Designs:}
Using the similar concept of generating Boolean logic using RTL, DTL, and TTL technology, we have designed several payload designs. For instance, XOR logic is often used as a payload for inverting a victim signal once a trigger condition is observed.
A 2:1 MUX logic is used for functional leakage. 
Fig. \ref{fig:rtl} (c) shows a payload implementation using XOR logic in RTL technology. We also show the use of a single transistor to construct a payload that causes a critical signal to stuck at logic high or low. For example the two payloads in Fig. \ref{fig:A2} (b) and (c) can be connected to a trigger to cause such conditional fault. The payload in (b) causes the \texttt{Victim\_Mod} to become low when \texttt{TRIG\_EN} becomes low and keeps the transistor off. The payload in (c) causes the \texttt{Victim\_Mod} to become low when \texttt{TRIG\_EN} becomes high and keeps the transistor on.

  \begin{figure}[b]
 	\centering
 	  \includegraphics[width=.8\columnwidth]{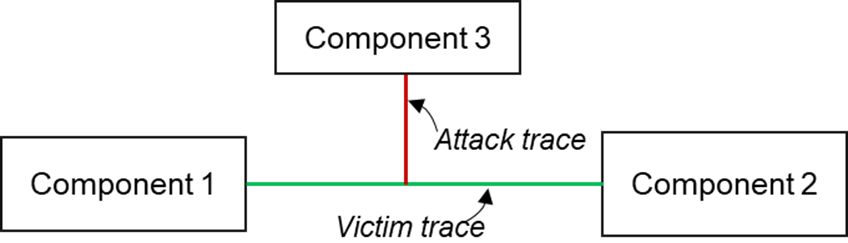}
 	\caption{Example of \textit{Alteration Model}: signal from component 3 with higher drive strength can overdrive the signal going from component 1 to component 2. Drive strength can be increased through the alteration of trace impedance or structure. }
 	\label{fig:alt}
 \end{figure}

If a design house is the victim and a foundry is the attacker, introducing such a Trojan is difficult due to the availability of a reference design to the design house for verification. However, inclusion of new components is still possible in a situation where the design house does not have the ability to observe the internal layers of the fabricated PCB through reverse engineering. Hence, the foundry could attempt to embed the additional components to the internal layers of the PCB \cite{ghosh2014secure}.

 \subsubsection{Alteration Model}
For the alteration model, we explore the various trace modification techniques presented in \cite{ghosh2014secure} and \cite{rosenfeld2010attacks}. Signal state hijacking techniques have been discussed in \cite{rosenfeld2010attacks} through the alteration of drive strengths of different signal connecting together. Such connections are common in JTAG or other debug infrastructures where the debug pins are connected to other pins of different board components. 
As shown in Fig. \ref{fig:alt}, a trace from component 1 is designed to drive component 3 and component 2. However, by configuring the pin of component 3 as an output with a higher drive strength than the output of component 1, the signal going to the component 2 can be controlled via component 3. Drive strength of a trace can be altered by modifying the trace impedance, length, thickness, or by integrating buffers \cite{rosenfeld2010attacks}.


\subsection{Sample PCB Designs for Trojan Insertion}
To generate the Trojan benchmarks, we take several sample, open-source PCB designs. More information about these sample designs can be found in \cite{git}. As shown in Table \ref{tab:sample}, these designs contain diverse components including microcontroller, FPGA, JTAG, transistors, resistor, capacitor, diode, etc. We used different trigger and payload structure combinations described in the addition model to create the Trojans and insert them in these sample PCB designs. To automate the insertion process, we used a Trojan insertion tool that will be described in the following section.

 \begin{figure}[!b]
 	\centering
 	  \includegraphics[width=.7\columnwidth]{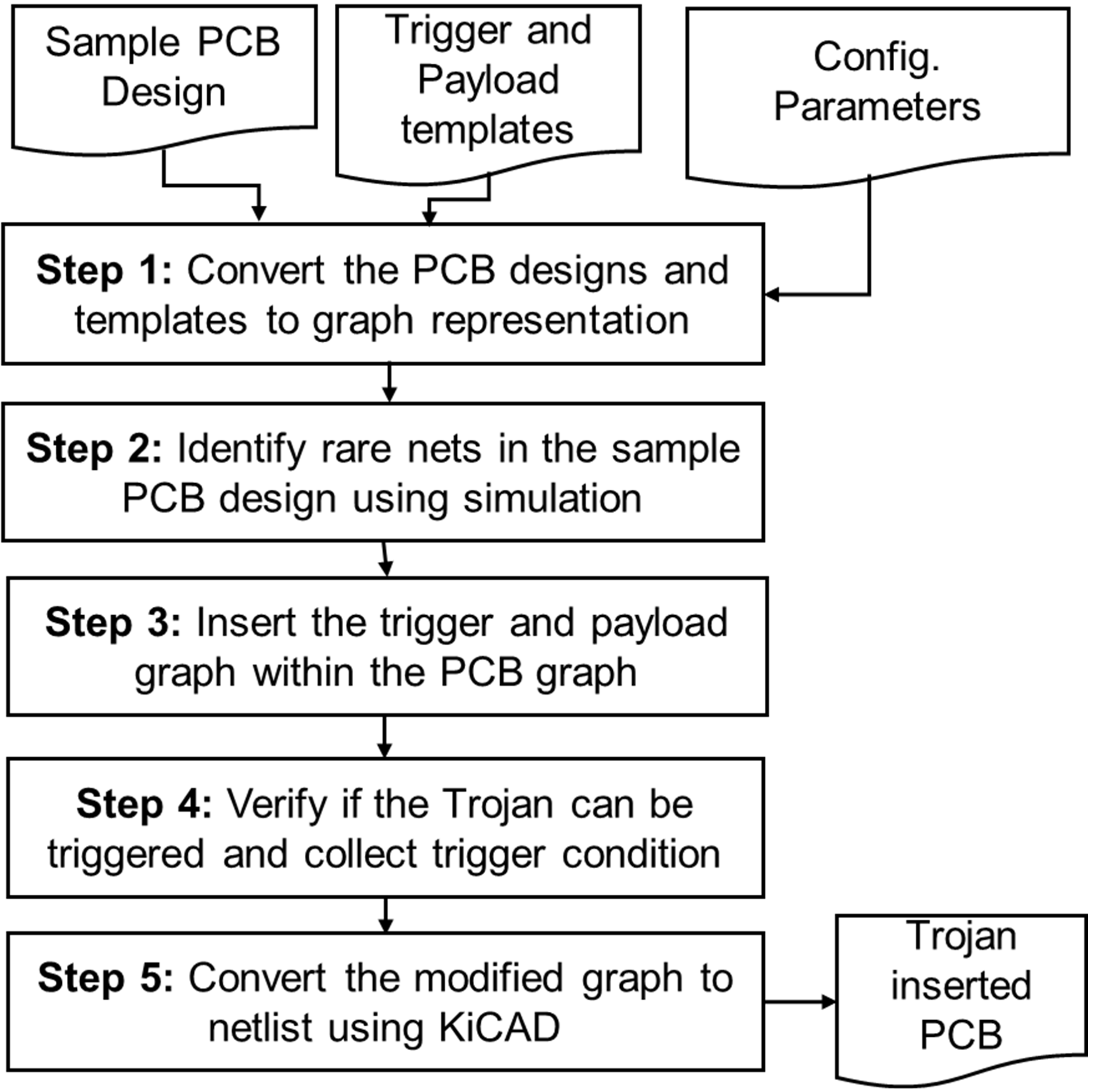}
 	\caption{Toolflow of the Trojan insertion tool.}
 	\label{TRIT}
 \end{figure}
\section{Automatic Tool-based Trojan Insertion}
\label{sec:synthetic}
 Our goal is to automatically generate a large number of Trojans with diverse structures that are hard to trigger. We utilize the existing trigger and payload designs in the addition model to generate different Trojans. The high-level methodology to automatically insert these Trojans is illustrated in Fig. \ref{TRIT}. 
The tool first takes user-provided configuration parameters which provide the option to select a desired trigger and payload circuit and the number of Trojans to be inserted. Based on this information, the tool selects the appropriate trigger and payload templates that are available in the form of a netlist. If the trigger and payload information is not specified by the user, the tool selects a random combination from default settings. The tool can also vary the value and arrangement of passive components (resistors, capacitors, inductors) of a given template to create structurally diverse, functionally equivalent templates. The passive component values and types are chosen based on the host PCB so as to not select a value vastly different than existing numbers.  

The tool takes the netlist representation of a sample PCB design as an input. To generate the netlist, we use KiCAD PCB design software \cite{kicad}. 
Since we are looking to insert hard-to-activate Trojans, the original PCB design can be simulated to find rarely switching traces. The appropriate traces that can be used as the input of a trigger must rarely reach the activation value of that trigger circuit. Otherwise, the activation condition may appear easily during functional verification by the consumer and the Trojan will get triggered. For simulation purposes we use external tools for spice simulation. If a CAD tool can store the simulation results of different traces in a file, our tool can utilize that. Proteus software can perform such board-level simulation \cite{wu2012application}. 
Our tool can analyze the simulation results to automatically find the appropriate input traces for connecting the trigger templates. It can also select from user-provided traces. Similarly, the victim signal to be leaked or corrupted can be specified by the user in the configuration file, or can be selected randomly by the tool.
   Next, the tool converts the sample PCB design to a graph structure, ($G_{PCB}$), to facilitate modification of the design. Graph representation of the trigger and payload template is augmented to the selected locations in $G_{PCB}$. The modified graph is then converted back to a PCB netlist containing the Trojan using the tool. The netlist can be read using KiCAD. The same process is repeated for generating the required number of Trojans specified by the user. 
 KiCAD provides the ability to automatically generate the layout.

\textbf{Benchmark Generation and Verification:} To test the toolflow, we provided the tool with the netlist of 5 different triggers and 3 different payloads. The tool inserted all possible combinations (15 total) of trigger and payload to each of the sample PCB designs and generated a total of 150 PCB Trojan benchmarks. To verify if these Trojans are inserted correctly, we performed two forms of verification: connectivity check and design rule violation check. For connectivity check, the tool reads back the Trojan inserted netlist and checks if there is a path between the location of Trojan insertion and the Trigger nets. For design rule violation check, the netlist files are read back with the KiCAD tool and automatically converted to layout. The KiCAD tool can identify possible design rule violations in the Trojan inserted design during this process. These PCB Trojan benchmarks will be made available to the public via \cite{TrojanBenchmarks}. The database may contain more Trojans as we improve the capabilities of the tool.

\begin{table}[t!]
\centering
\label{tab:fab}
\caption{Triggers and Payloads of the Fabricated Trojans}
\begin{tabular}{|l|l|l|}
\hline
\multicolumn{1}{|c|}{Trojan} & \multicolumn{1}{c|}{Trigger} & Payload        \\ \hline
Trojan1                           & 4 input AND                  & XOR                               \\ \hline
Trojan2                           & 4 input NOR                  & XOR                              \\ \hline
Trojan3                           & 8 input Mixed (AND + NOR)               & XOR                          \\ \hline
Trojan4                           & 8 input Mixed with Capacitor and diode & MUX                        \\ \hline
Trojan5                           & 8 input Mixed with Capacitor & MUX      \\ \hline
\end{tabular}
\end{table}

\section{Experimental Results from Fabricated PCB}
\label{sec:result}
\begin{figure}[t]
 	\centering
 	  \includegraphics[width=.9\columnwidth]{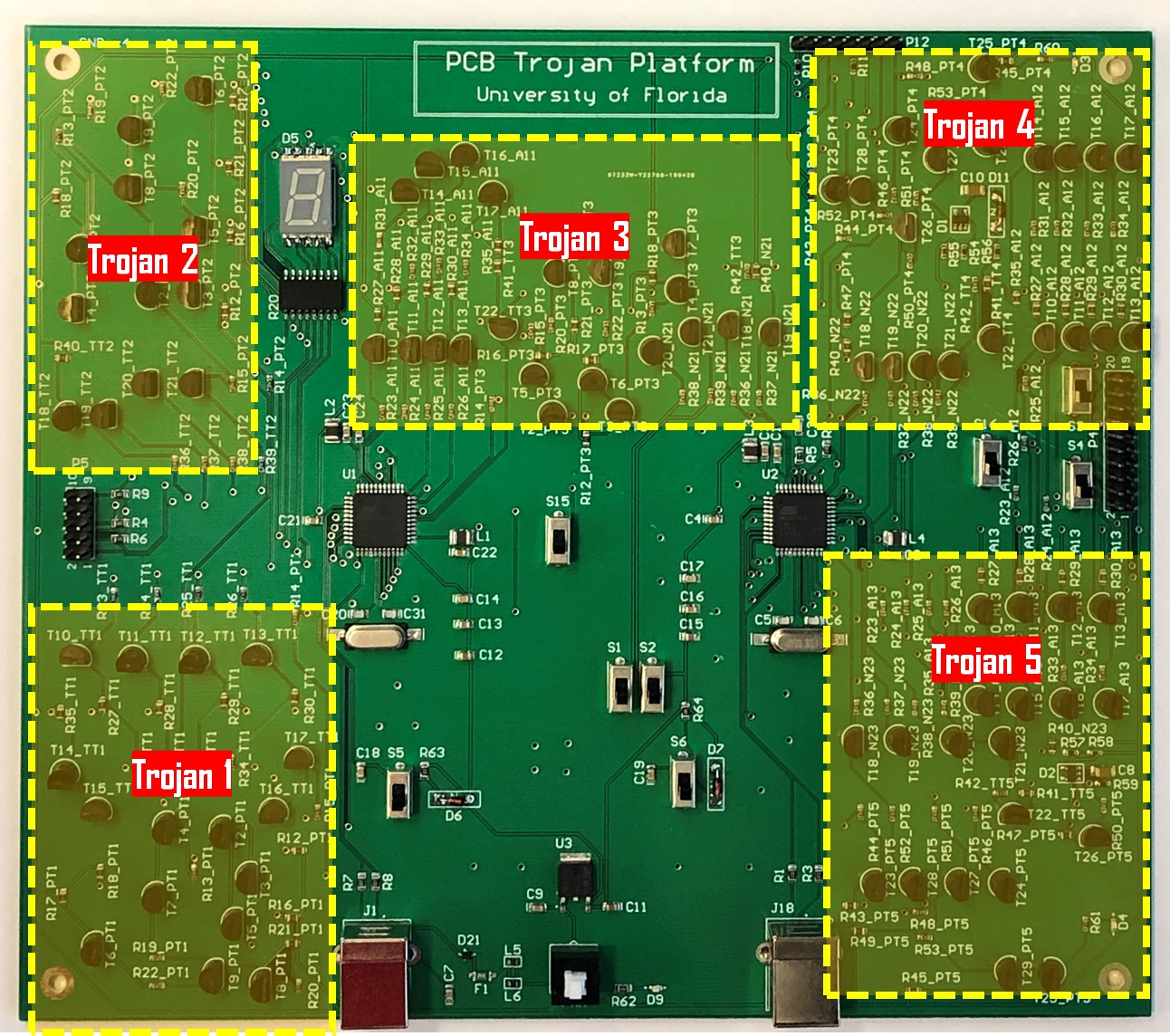}
 	\caption{Fabricated PCB Trojan platform with five Trojans.}
 	\label{fig:board}

 \end{figure}

We designed a PCB Trojan platform by integrating five Trojans with different trigger and payload combinations as listed in Table \ref{tab:fab}. We used two 8-bit AVR microcontrollers to drive all Trojan trigger inputs. The XOR payloads were connected such that they corrupt various victim signals going to seven segment displays. The MUX payloads were connected to leak victim signals via highly probable or observable points in the PCB, such as IOBs or LEDs. The fabricated PCB Trojan platform is shown in Fig. \ref{fig:board}.

 \begin{figure}[t]
 	\centering
 	  \includegraphics[width=.7\columnwidth]{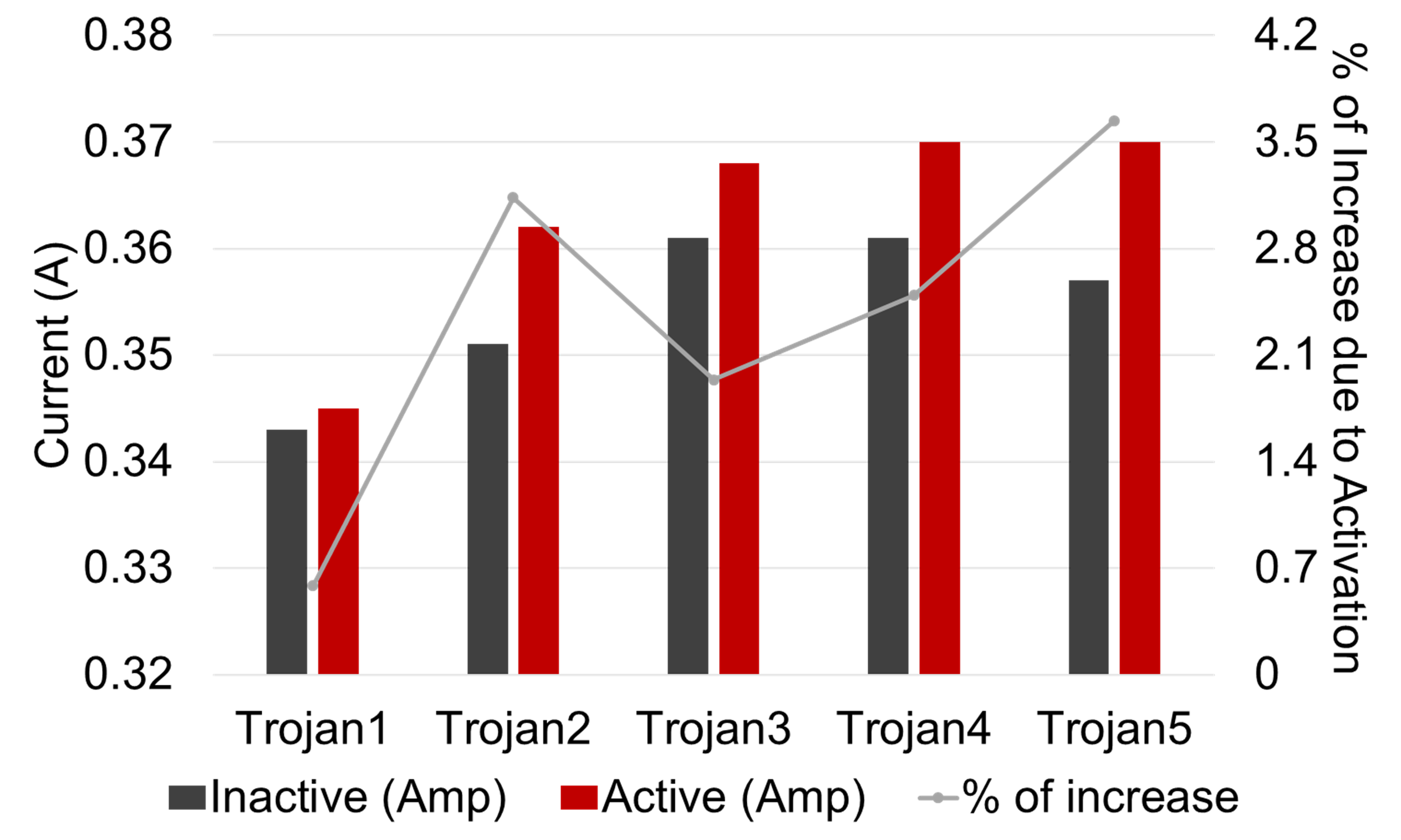}
 	\caption{Increase in current consumption due to Trojan activation (obtained from fabricated PCB).   }
 	\label{fig:current}
 	\end{figure}

To analyze the side-channel footprint of each Trojan, we measured the board-level current while keeping the Trojan in two states: inactive and active. In inactive mode, the input to the Trojan under consideration was kept fixed at a value that does not trigger the Trojan. Random vectors generated from the two microcontrollers were applied to the rest of the design that includes the other four Trojans. In active mode, we caused switching activity in the Trojan under consideration by applying vectors that alternatively activated and deactivated the Trojan in quick intervals. The activity of the rest of the PCB was kept the same as in the inactive mode. The current consumption of the five Trojans in active and inactive states is presented in Fig. \ref{fig:current}. We can observe a maximum of 3.5\% increase in current consumption (for Trojan5). Without any reference current signature to compare, such minor increase in current consumption is unlikely to raise any suspicion.      

\section{Conclusion and Future Work}
\label{sec:conclusion}
We have presented a taxonomy for board-level malicious circuits with associated designs of Trojans, specially for scenarios where the PCB design is untrusted. We also have presented a toolflow for generating large number of Trojan inserted PCB designs automatically. We analyzed the feasibility of the Trojan designs by fabricating a set of representative Trojans in a PCB and observing their side-channel footprint. Future work will include the development of always-on Trojans for PCBs with low area footprint and experimental evaluation of a number of Trojans from alteration model that does not require any incorporation of new components.

\ifCLASSOPTIONcaptionsoff
  \newpage
\fi



%
\bibliographystyle{IEEEtran}
\bibliography{IEEEexample}

%




\end{document}